\documentclass[a4paper,11pt]{article}
\usepackage{pos}
\usepackage{array}
\usepackage{multirow}
\usepackage{url}
\usepackage{color}
\usepackage{graphicx}
\usepackage{esint}
\usepackage{hyperref}
\usepackage{atbegshi}
\usepackage{lipsum}
\usepackage{booktabs}

\title{The spectral reconstruction of inclusive rates}
\ShortTitle{The spectral reconstruction of inclusive rates}

\author*[a]{John Bulava }

\affiliation[a]{Deutsches Elektronen-Synchrotron DESY, Platanenallee 6, 15738 Zeuthen, Germany}

\emailAdd{john.bulava@desy.de}

\abstract{A recently re-discovered variant of the Backus-Gilbert algorithm for 
spectral reconstruction enables the controlled determination of smeared 
spectral densities from lattice field theory correlation functions. 
A particular advantage of this approach is the \emph{a priori} specification of the kernel with which the underlying spectral density is smeared, allowing for variation of its peak position, smearing width, and functional form. If the unsmeared spectral density is sufficiently smooth in the neighborhood of a particular energy, 
it can be obtained from an extrapolation to zero smearing-kernel width at fixed peak position. 
	A natural application for this approach is scattering processes summed over all hadronic final states. As a proof-of-principle test, an inclusive rate is 
	computed in the 
	two-dimensional O(3) sigma model from a two-point correlation function of conserved currents. The results at finite and zero smearing radius 
	are in good agreement with the known analytic form up to energies at which 40-particle states contribute, and are sensitive to 
	the 4-particle contribution to the inclusive rate. 
	The straight-forward adaptation to compute the $R$-ratio in lattice QCD from two-point functions of the electromagnetic current is briefly discussed. }

\FullConference{%
The 39th International Symposium on Lattice Field Theory,\\
8th-13th August, 2022,\\
Rheinische Friedrich-Wilhelms-Universität Bonn, Bonn, Germany
}

\begin{document}
\maketitle
	
\section{Introduction}\label{s:intro}

Lattice QCD simulations proceed by computing $n$-point euclidean correlation functions of (quasi-) local interpolating operators. Single-hadron states and 
finite-volume few-hadron states are isolated from correlation functions in the asymptotic large euclidean time limit. However, some hadronic phenomena are best studied by other means. As an example, this work considers inclusive rates 
defined as a sum over all hadronic final states produced by an external current. At large center-of-mass energies, a finite-volume approach to such a process is impractical since it requires the isolation of all individual finite-volume levels with arbitrarily many particles. Such processes are a cornerstone of QCD and connect the low-energy hadronic and high-energy perturbative regimes~\cite{Pich:2020gzz}, serving as a manifestation of `quark-hadron duality'~\cite{Poggio:1975af} whereby perturbative QCD in terms of quarks and gluons becomes increasingly effective at computing inclusive rates summed over final states consisting 
entirely of hadrons. 

For concreteness, consider the QCD part of the process $e^+e^- \rightarrow\, {\rm hadrons}$
\begin{gather}\label{e:rrat}
\rho(s) = \frac{R(s)}{12\pi^2}, \quad R(s) = \frac{\sigma\left[{e}^+ {e}^- \rightarrow {\rm hadrons}\right](s)}{4\pi\alpha_{\rm em}(s)^2/(3s)},
\\ \rho_{\mu\nu}(k) = \frac{1}{2\pi}\int d^4 x \, e^{-ik\cdot x} \langle
\Omega | \hat{j}^{\rm em}_{\mu}(x)\, \hat{j}^{\rm em}_{\nu}(0) | \Omega \rangle = (g_{\mu\nu}k^2-k_\mu k_{\nu}) \, \rho(k^2),
\end{gather}
where $\hat{j}^{\rm em}_{\mu}$ is the quark-level electromagnetic current. The desired inclusive rate is given by the spectral density $\rho(s)$, which is also present in the analogous infinite-volume Euclidean correlator 
\begin{gather}\label{e:corr}
C(t) =  \int d^3 \boldsymbol{x} \, \langle \Omega | \hat{j}^{\rm em}_z(\boldsymbol{x}) \, e^{-\hat{H}t} \, \hat{j}^{\rm em}_z(0)^\dagger | \Omega \rangle = \int_0^{\infty} d\omega \, \omega^2 \rho(\omega^2) \, e^{-\omega t} \,.
\end{gather}
The direct determination of $\rho(s)$ in lattice QCD is not straightforward, however. First, the inversion of integral equations like Eq.~\ref{e:corr} using $C(t)$ evaluated at a finite number of discrete times with statistical errors is notoriously ill-posed. Furthermore, the finite volume introduces additional complication. Even if the inverse problem were solved successfully and the finite-volume euclidean correlator $C_{L}(t)$ used to determine its spectral density $\rho_{L}(s)$, it differs qualitatively from its infinite-volume counterpart $\rho(s)$. While $\rho_L(s)$ is a sum over Dirac $\delta$-functions for each finite-volume state, $\rho(s)$ is smooth apart from non-analyticities due to the opening of thresholds. In no way does $\rho_L(s)$ `approach' $\rho(s)$ as $L\rightarrow \infty$.  

The bridge between finite and infinite volume is made more effectively using the smeared spectral density 
\begin{align}
\label{e:rhosmear}
 \rho_{\epsilon}(E) = \int_0^\infty d\omega \, \delta_{\epsilon}(E-\omega) \, \rho(\omega) \,,
\end{align}
where  $\lim_{\epsilon \to 0} \delta_{\epsilon}(x) = \delta(x)$, $\delta(x)$ is the Dirac-delta function, and $\int_{-\infty}^\infty dx\, \delta_{\epsilon}(x) = 1$. This solves (in principle) both of the difficulties mentioned above: the inverse problem can be made arbitrarily mild by increasing in the smearing width $\epsilon$ 
and $\rho_{L, \epsilon}(E)$ approaches its infinite-volume counterpart in a 
well-defined manner. The goal is now to take the ordered double limit~\cite{Hansen:2017mnd} 
\begin{align}\label{e:dlim}
\rho(E) = \lim_{\epsilon \rightarrow 0^+} \lim_{L \rightarrow \infty} \rho_{L,\epsilon}(E), 
\end{align}
the asymptotic corrections to which are discussed in Sec.~\ref{s:o3}\footnote{Finite lattice spacing effects must also be removed by taking the continuum limit, which is here performed at fixed $\epsilon$ and $E$.}. 

Although spectral reconstruction has a long history in lattice QCD, particularly at finite temperature~\cite{Kaczmarek:2022ffn}, 
the treatment of the inverse problem in Eq.~\ref{e:corr} demands special care. In order to define the result of the spectral reconstruction procedure, precise knowledge of the smearing kernel in Eq.~\ref{e:rhosmear} is required. 
As detailed in Sec.~\ref{s:o3}, the Backus-Gilbert approach~\cite{BG1, BG2} is 
suitable in this respect\footnote{Another spectral reconstruction algorithm for which the smearing kernel is formally known \emph{a posteriori} is the Chebyshev polynomial approach of Ref.~\cite{Bailas:2020qmv}. Ref.~\cite{Barone:2022gkn} compares that approach to the one employed here. }. Since the estimator for the smeared spectral density $\hat{\rho}_{\epsilon}(E)$ is simply a linear combination of the input correlator data $\hat{\rho}_{\epsilon}(E) = \sum_t g_t(\epsilon, E) C(t)$, the resultant smearing kernel is given by the same linear combination of decaying exponentials in Eq.~\ref{e:corr}
\begin{align}\label{e:ker}
	\hat{\rho}_{\epsilon}(E) = \int d\omega \, \hat{\delta}_{\epsilon}(E,\omega)\, \rho(\omega), \qquad	\hat{\delta}_{\epsilon}(E,\omega) = \sum_t g_t(\epsilon,E) \omega^2 {e}^{-\omega t}.
\end{align}
Explicit knowledge of $\hat{\delta}_{\epsilon}(E,\omega)$ is a minimum requirement for a well-defined spectral reconstruction procedure. 

 Naively the Backus-Gilbert approach provides knowledge of the kernel in 
 Eq.~\ref{e:ker} only \emph{a posteriori} for a given choice of coefficients.  
 However, the coefficients themselves can be chosen to approximate a particular 
 smearing kernel specified \emph{a priori}~\cite{pt_hlt}. This important innovation is employed here and was applied to lattice field theory for the first 
time in Ref.~\cite{Hansen:2019idp}. In order to understand this reconstruction algorithm, and in particular demonstrate control over the systematic errors, 
a test in a controlled context is warranted. 
Such a test has been performed in Ref.~\cite{Bulava:2021fre} for the two-dimensional O(3) sigma model together with attempts at saturating 
the ordered double limit of Eq.~\ref{e:dlim}.

The remainder of this work is organized as follows. The spectral reconstruction method is presented in Sec.~\ref{s:o3} in the context of the  O(3) model test mentioned above. Prospects for adapting the method to current correlators in lattice QCD is discussed in Sec.~\ref{s:qcd} and Sec.~\ref{s:conc} concludes.

\section{O(3) model test}\label{s:o3}

\begin{figure}
	\includegraphics[width=0.49\textwidth]{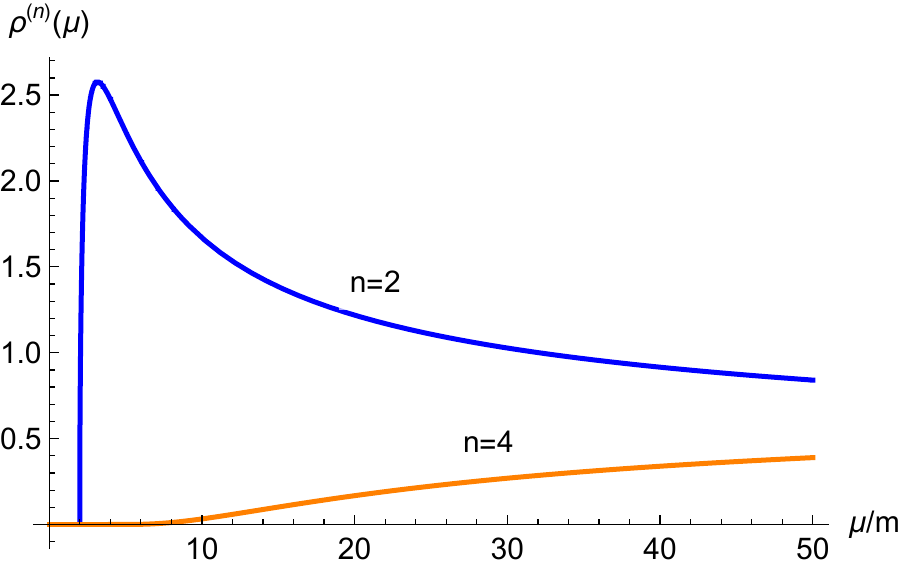}
	\includegraphics[width=0.49\textwidth]{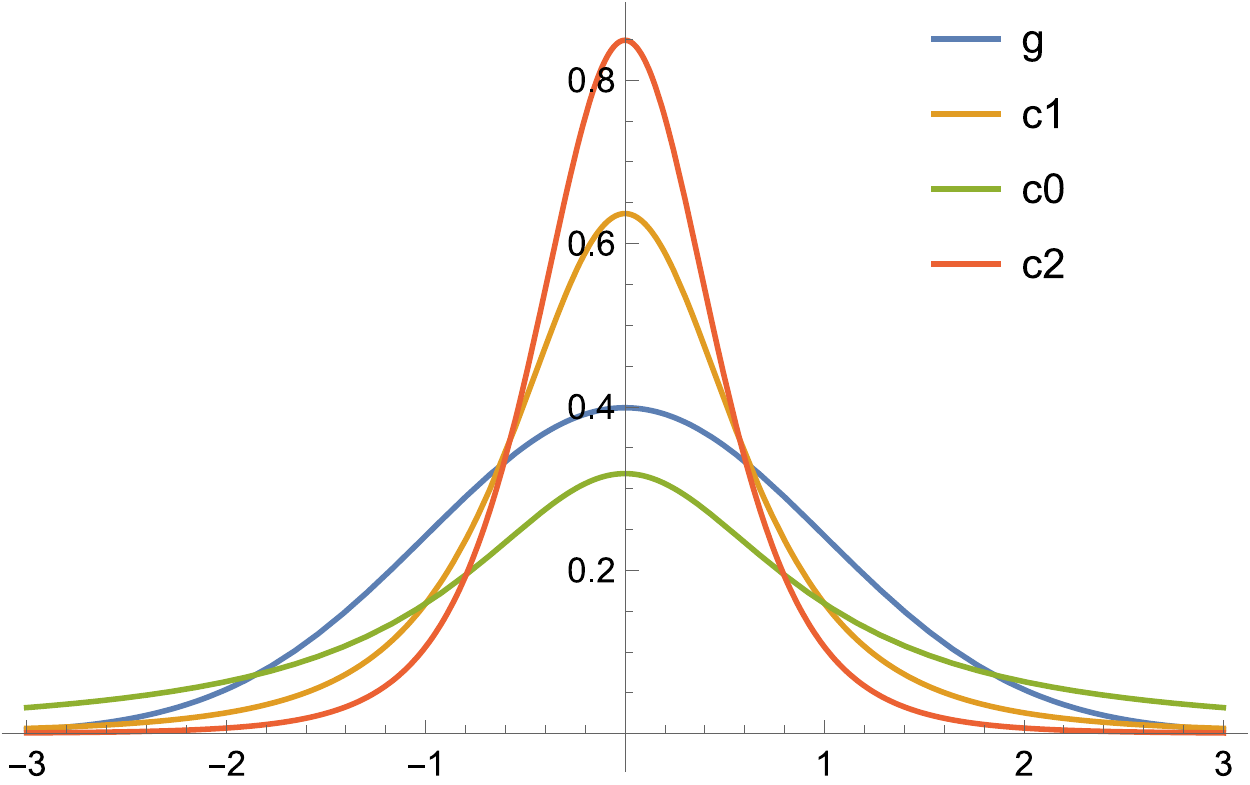}
	\caption{\label{f:rho_and_smear} {\bf Left}: exactly known $n=2$ and $n=4$ particle 
	contributions to the (continuum, infinite-volume) spectral density associated with the conserved current in the two-dimensional O(3) sigma model. Contributions from states with more particles are insignificant in the energy range shown here. {\bf Right}: the four smearing kernels $\delta^{\sf x}_{\epsilon}(x)$, where ${\sf x} = \{ {\sf g}, \, {\sf c}0, \, {\sf c}1, \, {\sf c}2\}$,
	defined in Eq.~\ref{e:kers}  plotted against $x$ with $\epsilon=1$ .} 
\end{figure}

This section reviews a spectral reconstruction test which was recently published in Ref.~\cite{Bulava:2021fre}. It employs the spectral reconstruction
procedure of Ref.~\cite{Hansen:2019idp} in the two-dimensional O(3) sigma model. Consider the standard lattice discretization 
\begin{align}\label{e:slat}
S[\sigma] = \frac{\beta}{2} \sum_{x \in \Lambda} a^2
\sum_\mu \hat{\partial}_{\mu} \sigma(x) \cdot \hat{\partial}_{\mu} \sigma(x)
=
\beta \sum_{x \in \Lambda}
\sum_\mu \left[ 1 - \sigma(x) \cdot \sigma(x + a\hat{\mu}) \right]
\,,
\end{align}
where $\sigma(x) \in \mathbb{R}^3$, $|\sigma(x)|=1$, and $\hat{\partial}_{\mu} f(x) = \frac{1}{a} [f(x + a\hat{\mu}) - f(x)]$. This model has a 
conserved current 
\begin{align}\label{e:lat_cur}
j_{\mu}^a(x) = \beta \epsilon^{abc} \sigma^b(x) \hat{\partial}_{\mu} \sigma^c(x)
\,
\end{align}
at finite lattice spacing, and possesses a dynamically-generated mass gap $m$. 
The total zero spatial momentum euclidean 
current-current correlation function analogous to Eq.~\ref{e:corr} (but without the factor of $\omega^2$) is computed by numerical simulations 
using the single-cluster algorithm of Ref.~\cite{Luscher:1990ck}.
A variety of ensembles are generated, with $mL \approx 30-60$ and 
$am \in [0.01, 0.04]$ to assess finite volume effects and take the continuum 
limit. 
In the continuum the contributions to the associated spectral density $\rho(\omega)$ from each fixed-particle number sector can be computed exactly~\cite{Balog:1996ey}. Below energies $E < 50m$, only the $n=2$ and $n=4$ particle 
contributions are significant and are shown in Fig.\ref{f:rho_and_smear}. 

In order to demonstrate the \emph{a priori} specification of the smearing 
kernel, consider four kernels with different profiles as a function of 
$x = E - \omega$:
\begin{align}\label{e:kers}
\delta_{\epsilon}^{\sf g}(x) & = \frac{1}{\sqrt{2\pi}\epsilon} \exp\left[-\frac{x^2}{2\epsilon^2}\right], &
\delta_{\epsilon}^{{\sf c}0}(x) & = \frac{1}{\pi} \frac{\epsilon}{x^2 + \epsilon^2} \,, \\[5pt]
\delta_{\epsilon}^{{\sf c}1}(x) & = \frac{2}{\pi} \frac{ \epsilon ^3}{ ( x^2 + \epsilon ^2 )^2} \,,
 & \delta_{\epsilon}^{{\sf c}2}(x) & = \frac{8}{3 \pi} \frac{ \epsilon ^5}{ ( x^2 + \epsilon ^2 )^3},
\,
\end{align}
including the gaussian (denoted `{\sf g}') and three Cauchy-like kernels denoted `{\sf c}$n$', for which $n = 0,1,2$ distinguishes the power of the pole. 
These kernels are depicted in Fig.~\ref{f:rho_and_smear}.

\begin{figure}
	\includegraphics[width=0.49\textwidth]{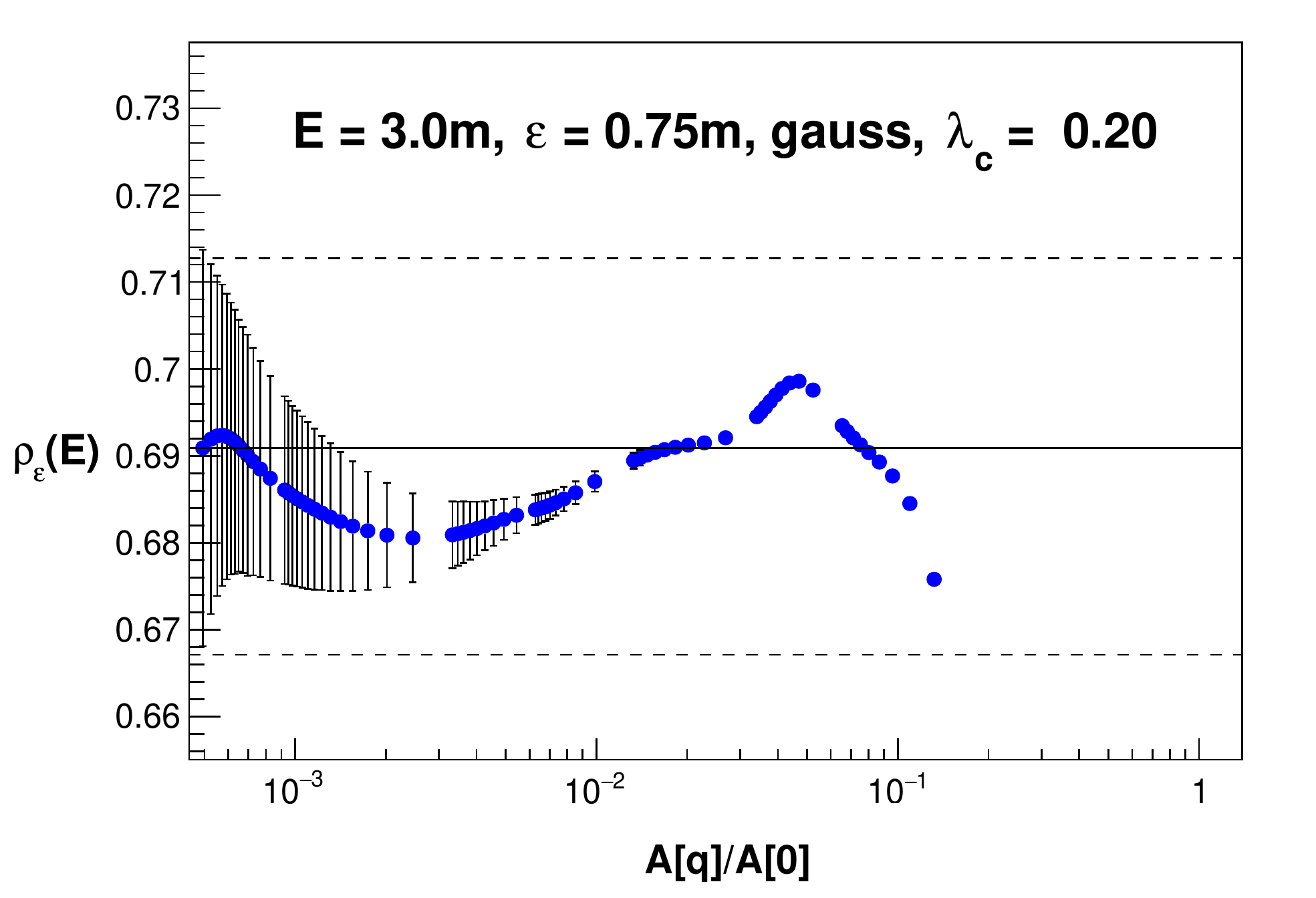}
	\includegraphics[width=0.49\textwidth]{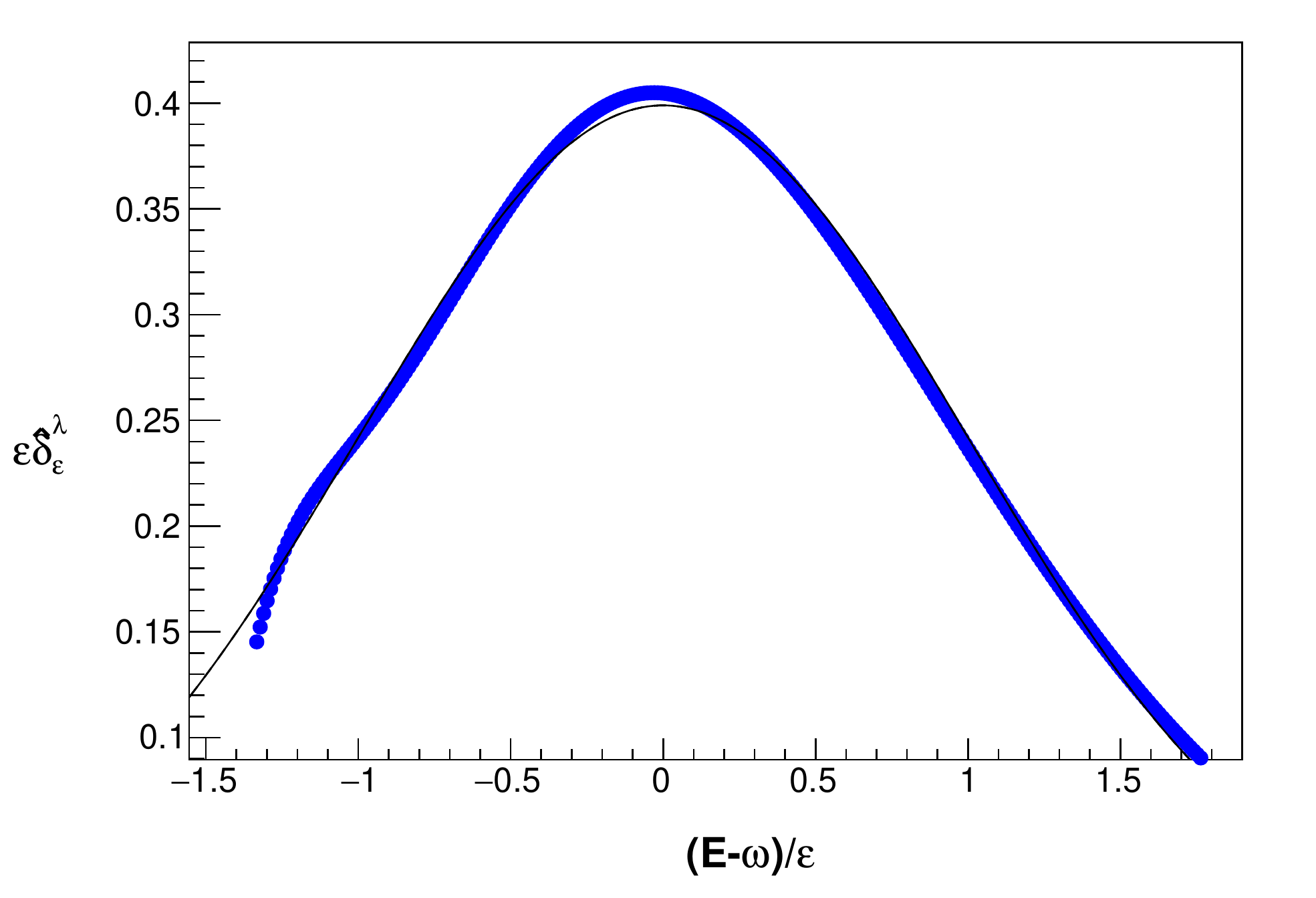}
	\caption{\label{f:recon} {\bf Left}: indicative illustration of the 
	trade-off between statistical and systematic errors for a particular choice of $E$ and $\epsilon$ on a single ensemble. Each point corresponds to a different 
	$\lambda$ and the horizontal band indicates the chosen reconstruction (with $\lambda = \lambda_{\rm c} = 0.20$) for which statistical errors dominate systematic ones. {\bf Right}: for this same setup and $\lambda = \lambda_{\rm c}$, the reconstructed kernel $\hat{\delta}^{\sf g}_{\epsilon}(E,\omega)$, together with the desired kernel $\delta_{\epsilon}^{\sf g}(E-\omega)$ shown as a solid line. 
	All reconstructions employ the correlator timeslices $t= 1a, \dots, 160a$.  }
\end{figure}

The method advocated in Ref.~\cite{Hansen:2019idp} to reconstruct the smeared spectral density $\rho^{\sf x}_{\epsilon}(E)$, where ${\sf x} = \{ {\sf g}, \, {\sf c}0, \, {\sf c}1, \, {\sf c}2\}$, is based on two criteria. First, the reconstructed smearing kernel $\hat{\delta}_{\epsilon}^{\sf x}(E,\omega)$ should be close to the desired one $\delta_{\epsilon}^{\sf x}(E-\omega)$. Second, the coefficients $\{ g_t(\epsilon, E) \}$ in Eq.~\ref{e:ker} should not induce a large
statistical variance on the estimator $\hat{\rho}^{\sf x}_{\epsilon}(E)$. These 
two considerations are encoded in the functionals 
\begin{gather}\label{e:funcs}
	A[g] = \int_{E_0}^{\infty} d\omega\, \left\{ \delta_{\epsilon}^{\sf x}(E-\omega) - \hat{\delta}_{\epsilon}^{\sf x}(E,\omega)\right\}^2, 
	\\
	B[g] = {\rm Var}[\hat{\rho}_{\epsilon}^{\sf x}(E)] =  
	\sum_{tt'} 
	g_t(\epsilon, E)\, 
	g_{t'}(\epsilon, E) \,{\rm Cov}[C(t), C(t')] 
\end{gather}
respectively. The coefficients are then chosen to minimize the combination functional $G_{\lambda}[g] = (1-\lambda)A[g]/A[0] + \lambda B[g]$, where the 
`trade-off' parameter $\lambda$ is introduced. For small $\lambda$,
the `accuracy' functional $A[g]$ takes preference over the `precision' one 
$B[g]$ resulting in small systematic but large statistical errors. By contrast, 
large $\lambda$ results in small statistical errors but a reconstructed smearing 
kernel which does not resemble the desired one. The choice of the parameter 
$\lambda$ is performed automatically and results in an approximate balance of 
these two criteria. The effect of varying $\lambda$ is illustrated in Fig.~\ref{f:recon} together with the reconstructed kernel $\hat{\delta}_{\epsilon}^{\sf g}(E, \omega)$ for a sample setup. The trade-off between statistical and systematic 
errors is familiar to lattice field theorists and the left panel of Fig.~\ref{f:recon} resembles the identification of a plateau in effective mass plots. 

\begin{figure}
  \includegraphics[width=\textwidth]{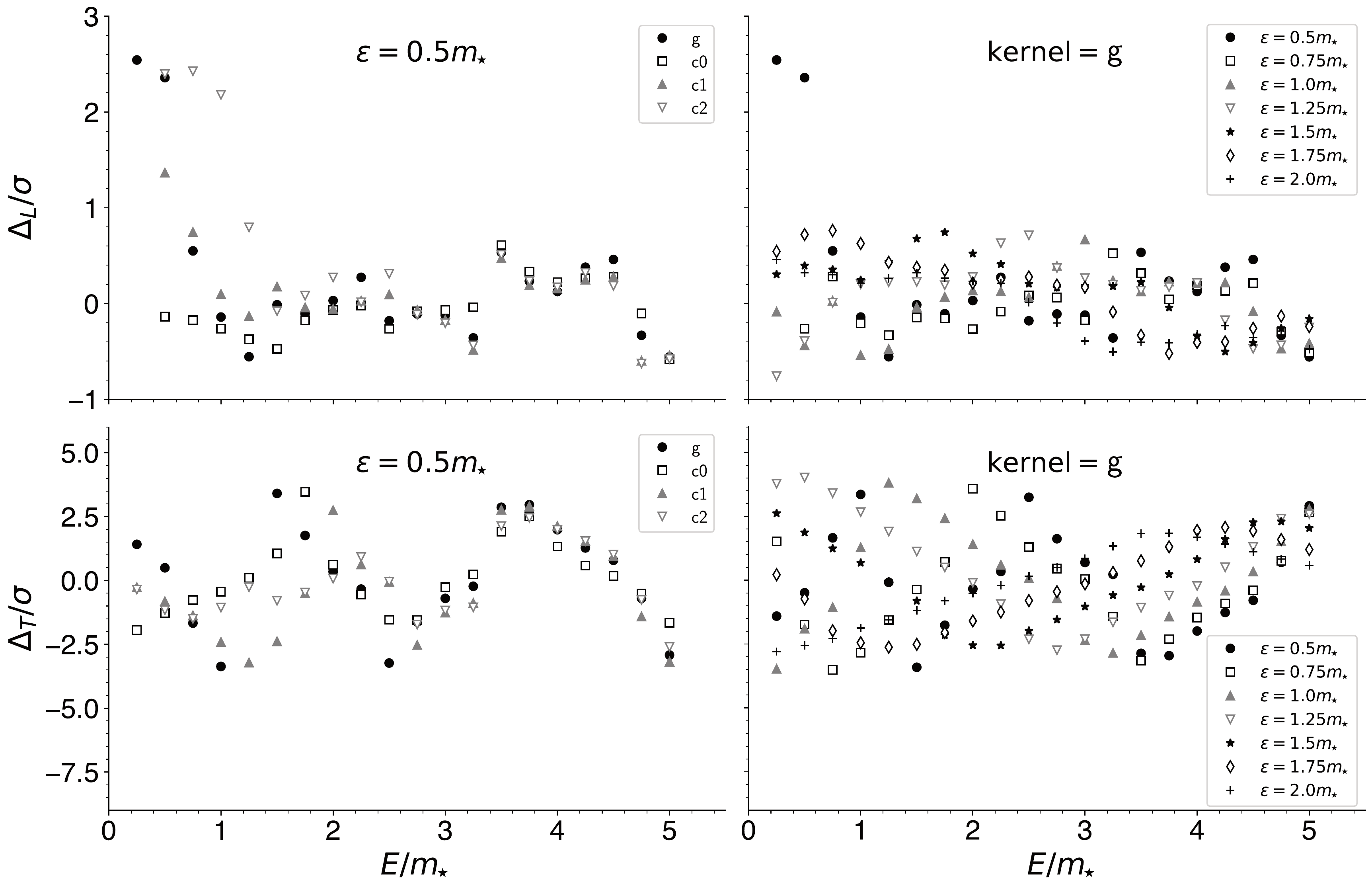}
	\caption{\label{f:lt}The difference $\Delta_{L,T}$  between spectral reconstruction on ensembles using $L$ and $2L$ (top row), and using ensembles with $T$ and $2T$ in the bottom row. In both cases $\Delta_{L,T}$ is divided by the statistical error on the smaller ensemble. Perhaps some marginally significant hints for 
	finite-$L$ effects are observed at small $E$ near the two-particle production 
	threshold. }
\end{figure}

The procedure described above is performed for a variety of $E$ and $\epsilon$, 
and for all four smearing kernels. Next, finite volume effects must be assessed 
and the continuum limit taken independently for each $E$, $\epsilon$, and kernel.
Finite-volume effects are assessed at a single lattice spacing by simulating two additional ensembles with doubled spatial and temporal extents, respectively. The differences $\Delta_{L,T}$ between the spectral reconstruction on the doubled lattices divided by the statistical error are shown in Fig.~\ref{f:lt}, which show (at most) moderately significant hints for finite-$L$ effects at energies near
the two-particle production threshold. 

\begin{figure}
	\includegraphics[width=\textwidth]{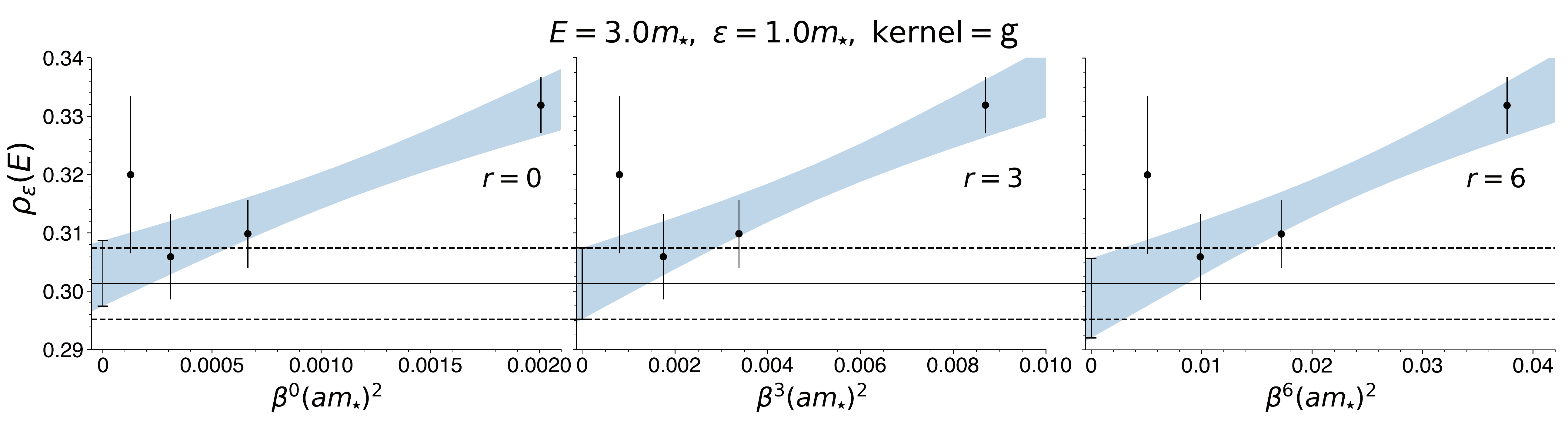}
	\caption{\label{f:cont}The continuum limit for a single $E$, $\epsilon$, and smearing kernel using the ansatz of Eq.~\ref{e:cont}. The shaded band indicates the fit, and the horizontal dotted region the extrapolated value for $r=3$, which
	is taken as the final result.}
\end{figure}

With the finite-volume effects demonstrably controlled for the ($E$, $\epsilon$) values and the kernels in question, the continuum limit can be investigated.
Cutoff effects for `on-shell' quantities in the two-dimensional O(3) sigma model have a long history, due to their apparently linear behaviour which is caused by large logarithmic corrections~\cite{Balog:2009np}. Unfortunately, the analysis there is 
incomplete for the `off-shell' smeared spectral density considered here. In order to explore the influence of logarithmic cutoff effects, the fit forms 
\begin{align}\label{e:cont}
	Q(a) = Q(0) + Ca^2\beta^r, \qquad r = 0,\,3,\,6
\end{align}
are explored to extrapolate the smeared spectral densities to the continuum 
limit. A comparison of the extrapolation forms is shown in Fig.~\ref{f:cont}.
The continuum limits are generally mild and well-constrained by the data, 
although the slope becomes steeper for increasing $E$.  

\begin{figure}
	\includegraphics[width=\textwidth]{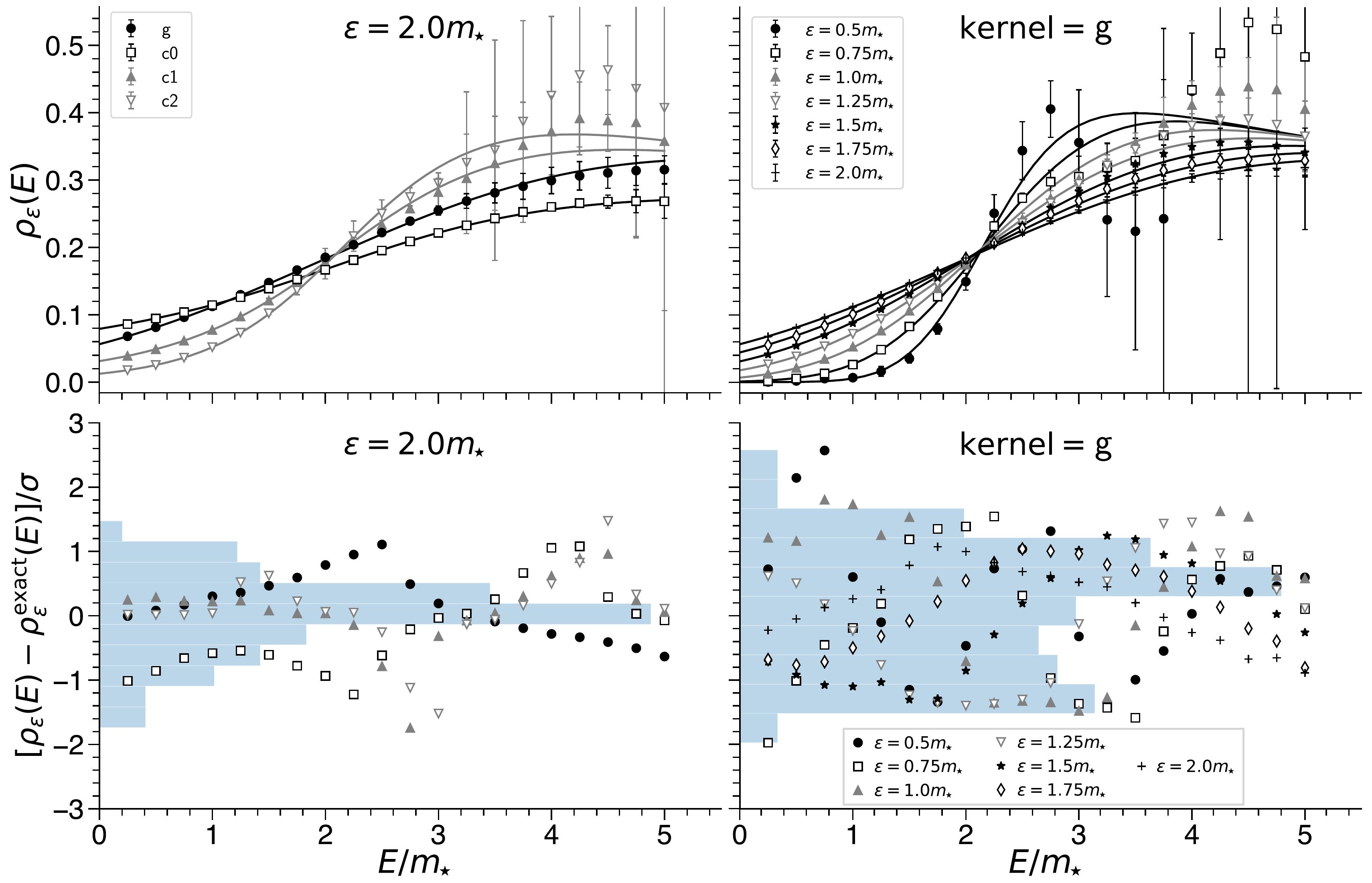}
	\caption{\label{f:sf} Lattice results for 
	$\rho_{\epsilon}^{\sf x}(E)$ in the continuum limit (the data points shown in the legend) compared against the exact spectral density including two-, four-, and six-particle contributions smeared with the exact kernel $\delta_{\epsilon}^{\sf x}(E-\omega)$. The exact results are shown as lines in the top row, and the bottom row shows the `pull' between the numerical data and the exact result, divided by the statistical and systematic error combined in quadrature. A naive 
	histogram of the differences, which ignores correlations among the data, 
	is shown horizontally in the bottom row and approximately resembles the unit gaussian. }
\end{figure}

Given the assessment of systematic errors due to spectral reconstruction, finite $L$ and $T$ effects, and finite lattice spacing, it is time to confront the computations of $\rho_{\epsilon}^{\sf x}(E)$ with the exact spectral density $\rho(\omega)$ (comprised of the two-, four-, and six-particle contributions) 
smeared with the exact smearing kernel 
$\delta^{\sf x}_{\epsilon}(E-\omega)$. Such a comparison is performed in Fig.~\ref{f:sf} where the numerical computations are demonstrably consistent with the exact results, within the quoted errors. These errors take into account the statistical errors due to the reconstruction, with any residual systematic errors 
from finite-$L,T$ effects and the continuum limit added in quadrature.  

At this point the verification of the spectral reconstruction approach of 
Ref.~\cite{Hansen:2019idp} is complete. Smeared spectral densities in the two-dimensional O(3) sigma model have been reconstructed with smearing kernels specified \emph{a priori}, which are consistent with the exact result after the continuum 
limit has been taken. Consider now the $\epsilon\rightarrow 0$ limit in Eq.~\ref{e:dlim}. For this, an important property of the unsmeared spectral density evident in Fig.~\ref{f:rho_and_smear} is required, namely that it varies increasingly 
slowly with increasing $E$. This circumvents the limitations in reconstructing 
kernels with a fixed $\epsilon$ and increasing $E$ evident in Fig.~\ref{f:sf}: 
larger smearing widths are sufficient at larger $E$. To this end, the smearing width is scaled $\epsilon \propto (E-2m)$. Also, rather than using a single smearing kernel, $\rho_{\epsilon}^{\sf x}(E)$ for all kernels are used to perform 
constrained extrapolations. For this the small-$\epsilon$ expansion is useful
\begin{align}
\label{e:expand}
\rho^{{\sf{x}}}_{\epsilon}(E) & \equiv \int_0^\infty d \omega \, \delta_{\epsilon}^{\sf x}(E - \omega) \, \rho(\omega) = \rho(E) + \sum_{k=1}^\infty w^{{\sf x}}_k a_k(E) \epsilon^{k} \,,
\end{align}
where the contribution at the $k$th order in $\epsilon$ is the product of a 
kernel-independent factor
\begin{equation}
\label{e:ak}
a_{k}(E) =
\begin{cases}
\frac{(-1)^{k/2}}{k!} \left( \frac{d}{dE} \right)^{k}\rho(E)\,,
\qquad & \textrm {$k$ even} 
\\
\\
\lim_{\eta \to 0^+} \frac{(-1)^{(k-1)/2}}{2\pi} \int_{-\infty}^\infty {\rm d}\omega \, \frac{\rho(E+\omega) + \rho(E-\omega)}{(\omega+i\eta)^{k+1}}\,,
\qquad & \textrm{$k$ odd}
\end{cases}
\ .
\end{equation}
 which depends on the unsmeared spectral 
density, and a kernel-independent piece $w_{k}^{({\sf x})}$ which is however 
independent of $\rho(\omega)$. The  $w_{k}^{({\sf x})}$ for the kernels used here
are given for all orders in Tab.~\ref{t:wcoeffs}. The {\sf c0} kernel is 
however not practically useful in such extrapolations due to the ${\rm O}(\epsilon)$ term. 
\begin{table}
\renewcommand{\arraystretch}{1.7}
\begin{tabular}{c c c || c c c c}
\toprule
\ \ \ {\sf x} \ \ \ & \ \ \ $w^{{\sf x}}_k,$ even\ $k$ \ \ \ & \ \ \ $w^{{\sf x}}_k,$ odd\ $k$ \ \ \ & \ \ \ $\phantom{+} w^{{\sf x}}_1$ \ \ \ & \ \ \ $\phantom{+} w^{{\sf x}}_2$ \ \ \ & \ \ \ $\phantom{+} w^{{\sf x}}_3$ \ \ \ & \ \ \ $\phantom{+} w^{{\sf x}}_4$ \ \ \
\\ \midrule
{\sf g} & $\displaystyle \frac{k!}{(-2)^{k/2} (k/2)!}$ & $0$ & $\phantom{+} 0$ & $-1$ & $\phantom{+} 0$ & $\phantom{+} 3$ \\
{\sf c0} & $\displaystyle \phantom{\frac{1}{2}} 1 \phantom{\frac{1}{2}} $ & $\displaystyle \phantom{\frac{1}{2}} 1 \phantom{\frac{1}{2}} $ & $\phantom{+} 1$ & $\phantom{+} 1$ & $\phantom{+} 1$ & $\phantom{+} 1$
\\
{\sf c1} & $\phantom{\frac{1}{2}} (1-k) \phantom{\frac{1}{2}}$ & \ \ $(1-k)$ \ \ & $\phantom{+} 0$ & $-1$ & $-2$ & $-3$
\\
{\sf c2} & \ \ $\displaystyle \frac{1}{3}(k-3)(k-1)$ \ \ & \ $\displaystyle \frac{1}{3}(k-3)(k-1)$ \ & $\phantom{+} 0$ & $-1/3$ & $\phantom{+} 0$ & $\phantom{+}1$ \\
\bottomrule
\end{tabular}
	\caption{The kernel-dependent coefficients $w^{{\sf x}}_k$ appearing in the small-$\epsilon$ expansion of Eq.~\eqref{e:expand}. For the {\sf c1} and {\sf c2} kernels, $w_3^{{\sf c1}}$ and $w_5^{{\sf c2}}$ (respectively) are the non-zero coefficients with lowest odd order. \label{t:wcoeffs}}
\renewcommand{\arraystretch}{1.0}
\end{table}

\begin{figure}
	\includegraphics[width=\textwidth]{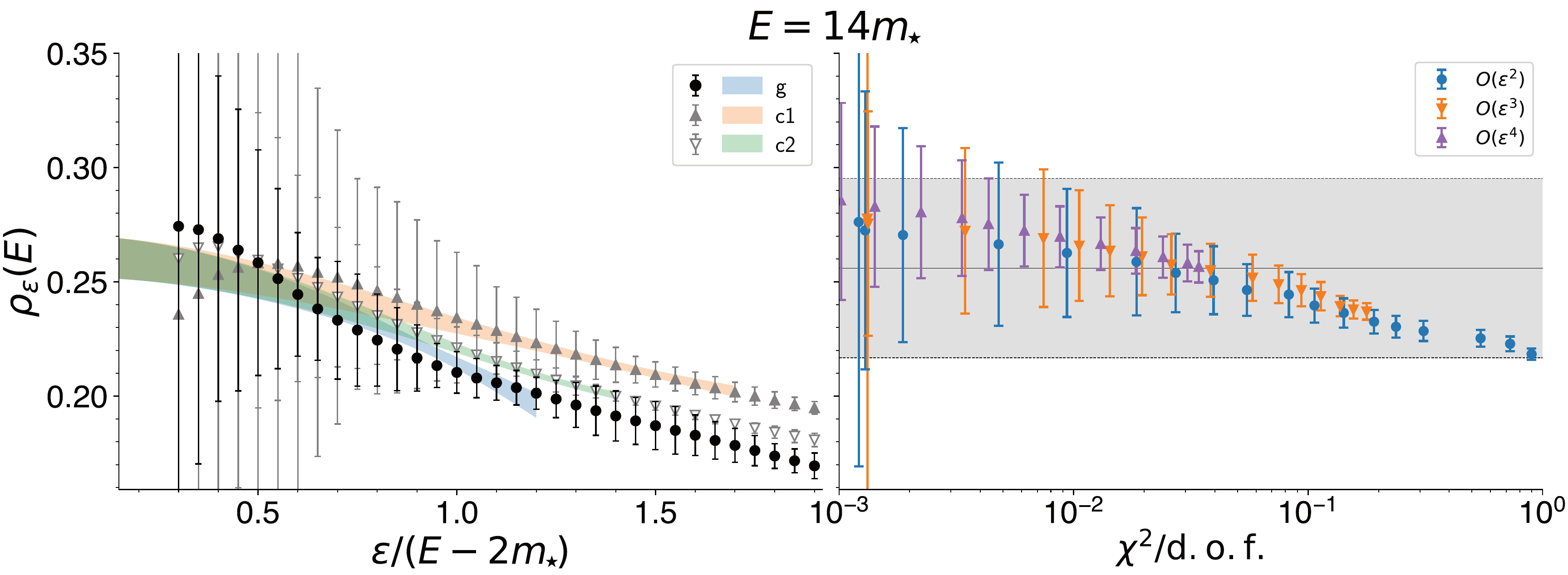}
	\caption{\label{f:eps_extrap} {\bf Left}: a sample constrained extrapolation using the known coefficients in Tab.~\ref{t:wcoeffs} up to and including ${\rm O}(\epsilon^4)$ terms for a fixed energy $E=14m$. The relative fit ranges of the different smearing kernels are adjusted so that each kernel has an equal amount of support between the two-particle threshold $2m$ and $E$. {\bf Right}: variation of the extrapolated value of $\rho(E)$ for different extrapolation ranges and orders. The final result is conservatively taken as the horizontal shaded region. } 
\end{figure}

A representative constrained extrapolation, in which all kernels (apart from 
{\sf c0}) are used to fit for $\rho(E)$ and the $a_k(E)$ up to a certain order, is shown
in Fig.~\ref{f:eps_extrap}. A final estimate for $\rho(E)$ is chosen with a statistical error larger than the variation between different extrapolation orders and ranges.    
Repeating this procedure for all values of $E$ yields the final results for the 
spectral density $\rho(E)$ shown in Fig.~\ref{f:final}. Not only do the numerical results agree with the exact spectral density including two-, four-, and six-particle contributions, but differ significantly from the two-particle contribution alone, indicating the sensitivity to four-particle states. Furthermore, the largest energy of $E=40m$ is statistically consistent with the two-loop perturbative 
result, demonstrating that $\rho(E)$ has been computed up to the onset of the perturbative regime.
\begin{figure}
	\includegraphics[width=\textwidth]{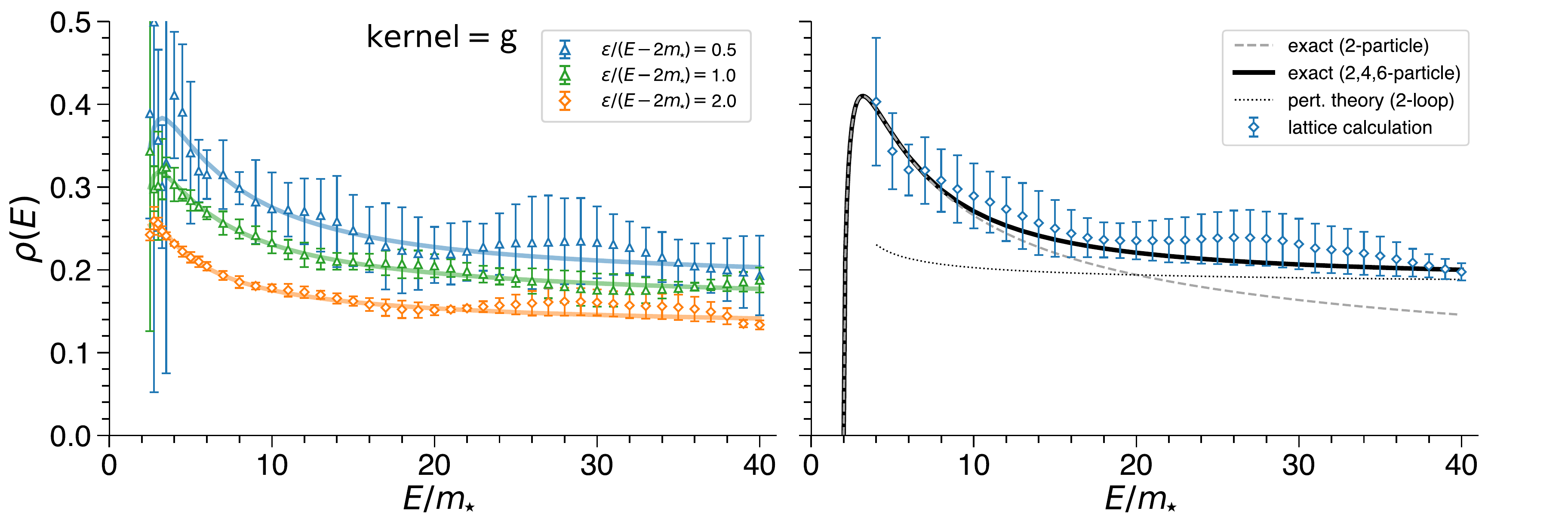}
	\caption{\label{f:final}{\bf Left}: a selection of some of values of $\epsilon$ (given in the legend) used in the $\epsilon\rightarrow 0$ extrapolation, together with the 
	exact smeared spectral density shown as solid lines for the gaussian kernel. 
	{\bf Right}: the final extrapolated results for $\rho(E)$ together with the 
	exact two-particle contribution to the spectral density and the sum of the two-, four-, and six-particle contributions. The two-loop perturbative spectral 
	density is also shown.}  
\end{figure}

\section{Prospects for QCD}\label{s:qcd}

It is in principle straightforward to adopt the analysis of the O(3) sigma model in Sec.~\ref{s:o3} to the lattice QCD computation of current spectral densities. 
However, while it is difficult to compare the density of finite-volume states in one and three spatial dimensions, the O(3) model setup 
with $mL \approx 30$ may be difficult to achieve in QCD. Fortunately, the 
masterfield paradigm~\cite{Luscher:2017cjh,Giusti:2018cmp,Francis:2019muy} 
offers the possibility of large lattice volumes by accumulating statistics from 
widely-separated space-time regions rather than widely-separated Markov chain elements. 

Work in this direction has been detailed at this conference in talks by 
M. C\`{e} and P. Fritzsch. This section describes preliminary work toward the spectral reconstruction of the isovector vector current spectral density 
with the collaborators and setup mentioned in those talks. 
Using $N_{\rm f} = 2+1$ dynamical flavors of stabilized Wilson fermions~\cite{Francis:2019muy} at $a=0.09\,{\rm fm}$, two ensembles were generated with $(L/a)^4 = 96^4$ and $192^4$. The analysis described below is based on two and five thermalized, widely separated configuations on the $L/a=96$ and $192$ ensembles, respectively. Details about the construction of the correlators and the estimation of the statistical errors were given by M. C\`{e}. For the data presented here, a variant of the bootstrap procedure is employed.  

As suggested in Sec.~\ref{s:intro}, the prototypical QCD 
analogue of Sec.~\ref{s:o3} is the 
hadronic component to $e^+e^- \rightarrow {\rm hadrons}$, which can be obtained 
by solving the inverse problem of the euclidean current-current correlator 
projected onto zero spatial momentum in Eq.~\ref{e:corr}. However, including 
both the isoscalar and isovector components of the electromagnetic current 
requires valence quark-line disconnected Wick contractions, incurring  
additional computational cost and statistical variance. Consider then 
the simpler case of the isovector-vector correlator. Phenomenologically this 
spectral density can be accessed directly from hadronic decays of the tau 
lepton~\cite{Davier:2005xq}. A state-of-the-art phenomenological 
determination of the isovector-vector spectral density is performed in Ref.~\cite{Boito:2021gqq}.

The spectral reconstruction approach of Sec.~\ref{s:o3} is adopted nearly 
identically here, apart from some key differences. First, the basis functions provided by the correlator data in Sec.~\ref{s:o3} are $b_t(\omega) = {\rm e}^{-\omega t} + {\rm e}^{-\omega(T-t)}$, but those employed in this analysis from Eq.~\ref{e:corr} are $b_t(\omega) = \omega^2 \, {\rm e}^{-\omega t}$. The flexibility of the
formalism of Sec.~\ref{s:o3} to handle these different basis functions is an 
advantage over the Chebyshev approach of Ref.~\cite{Bailas:2020qmv}. Also, for these large lattices the finite temporal extent can be demonstrably ignored. For a first test of the approach in QCD, only the gaussian smearing kernel from Eq.~\ref{e:kers} is considered. All correlator timeslices from $t_{\rm min} = a$ to $t_{\rm max} = 35a$ 
are used in the reconstruction, and all arithmetic operations are 
performed with 400 bits of computer precision using the \texttt{Arb} 
library~\cite{Johansson2017arb}. 

Another innovation for this analysis compared to Sec.~\ref{s:o3} is the procedure for choosing the $\lambda$ at which statistical errors dominate the systematic 
errors. As suggested by the left panel in Fig.~\ref{f:recon}, the procedure in Sec.~\ref{s:o3} which 
balances the two functionals $A[g]/A[0]$ and $B[g]$ from Eq.~\ref{e:funcs} is perhaps 
over-conservative and somewhat arbitrary. The alternative approach employed here makes use of one of the possible constraints introduced in Ref.~\cite{Bulava:2021fre}. By the addition of a lagrange multiplier, it is possible to 
enforce constraints on the reconstructed smearing kernel $\hat{\delta}_{\epsilon}^{\sf g}(E,\omega)$. Ref.~\cite{Bulava:2021fre} describes how to impose the coincidence of the reconstructed and desired kernels at a particular point
\begin{gather}
	\hat{\delta}^{\sf g}_{\epsilon}(E,\omega^*) = \delta^{\sf g}_{\epsilon}(E-\omega^*).
\end{gather}
Although Ref.~\cite{Bulava:2021fre} only considers $\omega = E$, the generalization to arbitrary $\omega^*$, even outside the interval $[E_0, \infty)$, is straightforward.

\begin{figure}
	\includegraphics[width=0.49\textwidth]{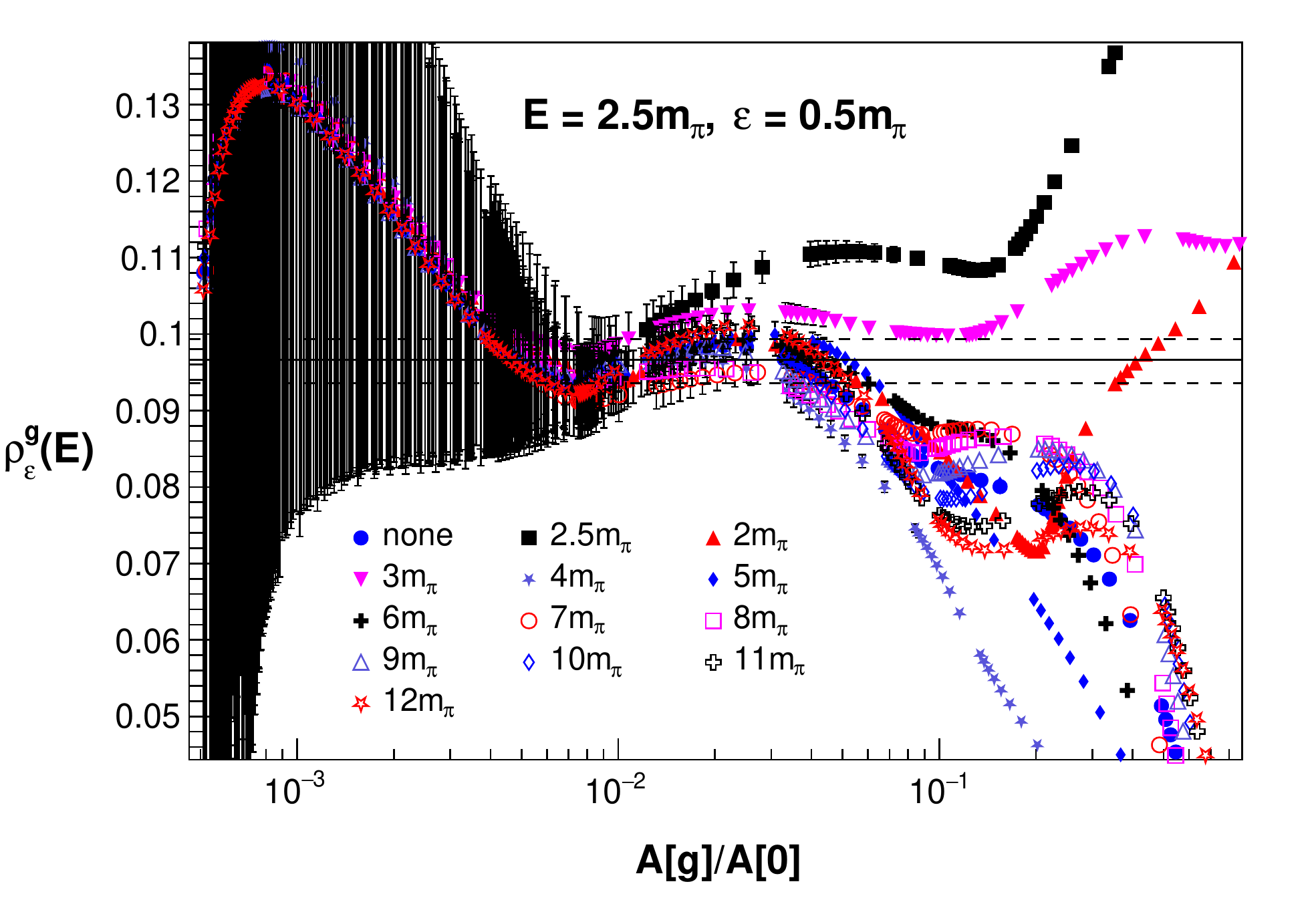}
	\includegraphics[width=0.49\textwidth]{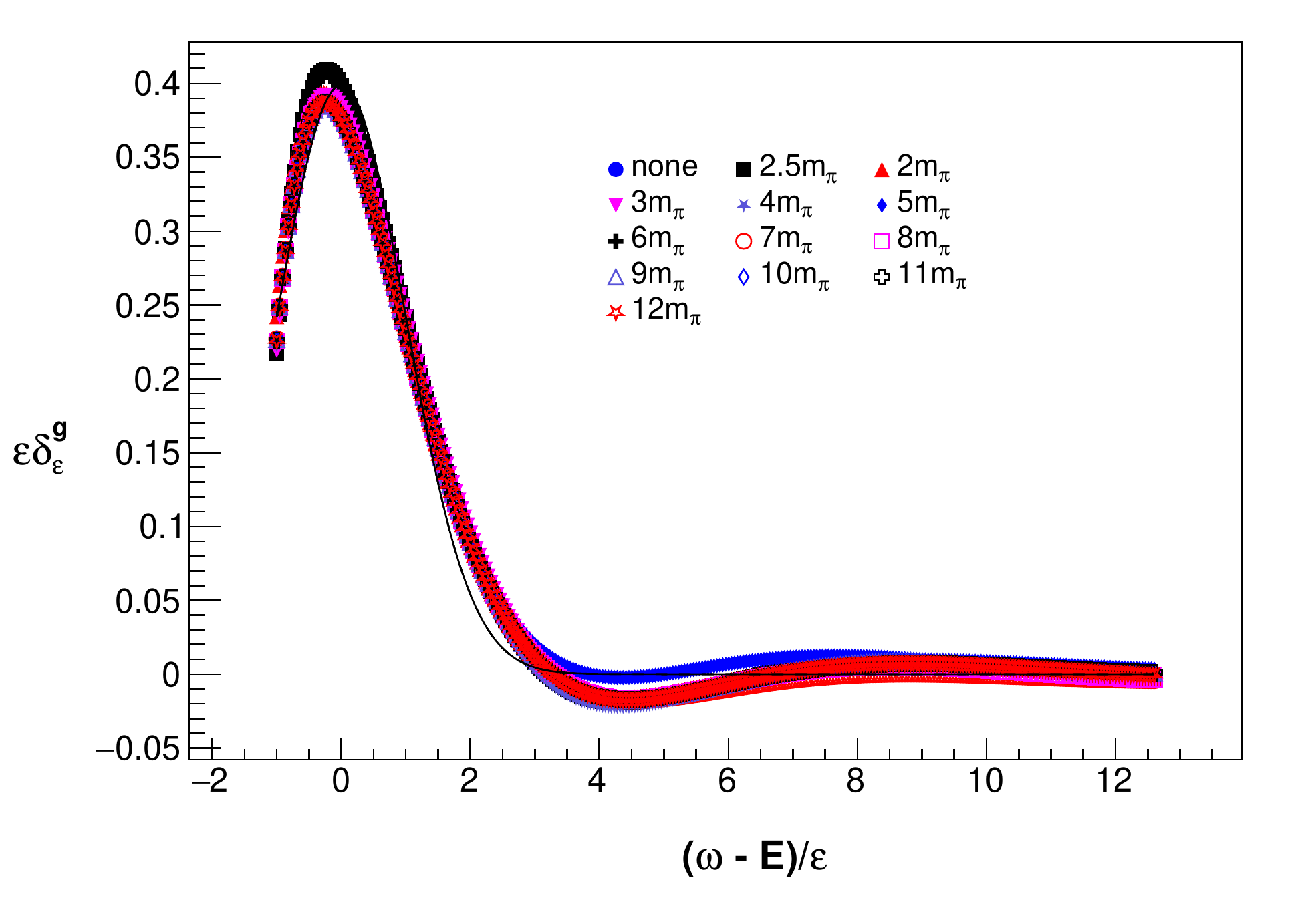}
	\caption{\label{f:const} Illustration of the method for choosing the optimal tradeoff parameter $\lambda$ described in the text for the gaussian reconstruction on the $L=18 \, {\rm fm}$ ensemble with $\epsilon = 0.5m_{\pi}$ and $E=2.5m_{\pi}$. {\bf Left}: different values of $\lambda$ for the unconstrained reconstruction and reconstructed kernels constrained to agree with $\delta^{\sf g}_\epsilon(E-\omega^*)$ at the various values of $\omega^*$ indicated in the legend. The horizontal band indicates the chosen estimate for which the statistical error on the unconstrained reconstruction covers the spread given by the ensemble of constraints. For comparison, the method for balancing statistical and systematic 
	errors of Sec.~\ref{s:o3} (and Ref.~\cite{Bulava:2021fre}) chooses the 
	unconstrained point 
	with $A[g]/A[0] \approx 0.0016$. 
	{\bf Right}: the reconstructed smearing kernel compared to the desired 
	gaussian (solid line) for each member of the constraint ensemble near the chosen value of $A[g]/A[0]$ indicated by the horizontal band in the left plot. The residual variation between the different constraints is evidently smaller than the statistical error on the constrained reconstruction, although perhaps additional values of $\omega^*$ near $\omega^*-E \approx 2\epsilon$ should be added in the future. }
\end{figure}

Using this `equal value' constraint on the reconstructed kernel, it is possible 
to estimate how small $A[g]/A[0]$ must be for the statistical errors to dominate. An `ensemble' of reconstructions are performed with different values of $\omega^*$, in addition to the unconstrained one. The systematic error estimate is then obtained from the variation of $\hat{\rho}^{\sf g}_{\epsilon}(E)$ among this ensemble at similar $A[g]/A[0]$. The point at which this variation is smaller than the statistical 
error on the unconstrained result is taken as the optimal reconstruction. 
Of course this procedure depends on the ensemble of constraint points $\{\omega^*\}$ which are considered. However, it is sensitive the 
unsmeared spectral density $\rho(\omega)$, in contrast to the approach of Ref.~\cite{Bulava:2021fre}. If additional values of $\omega^*$ are added for which $\rho(\omega^*)$ has little support, these will likely differ little from the unconstrained case, apart from possible variations in  $\hat{\delta}^{\sf g}_{\epsilon}(E,\omega)$
away from $\omega^*$ induced by the constraint at $\omega^*$. An illustration of this procedure is given in Fig.~\ref{f:const}.

\begin{figure}
	\includegraphics[width=0.49\textwidth]{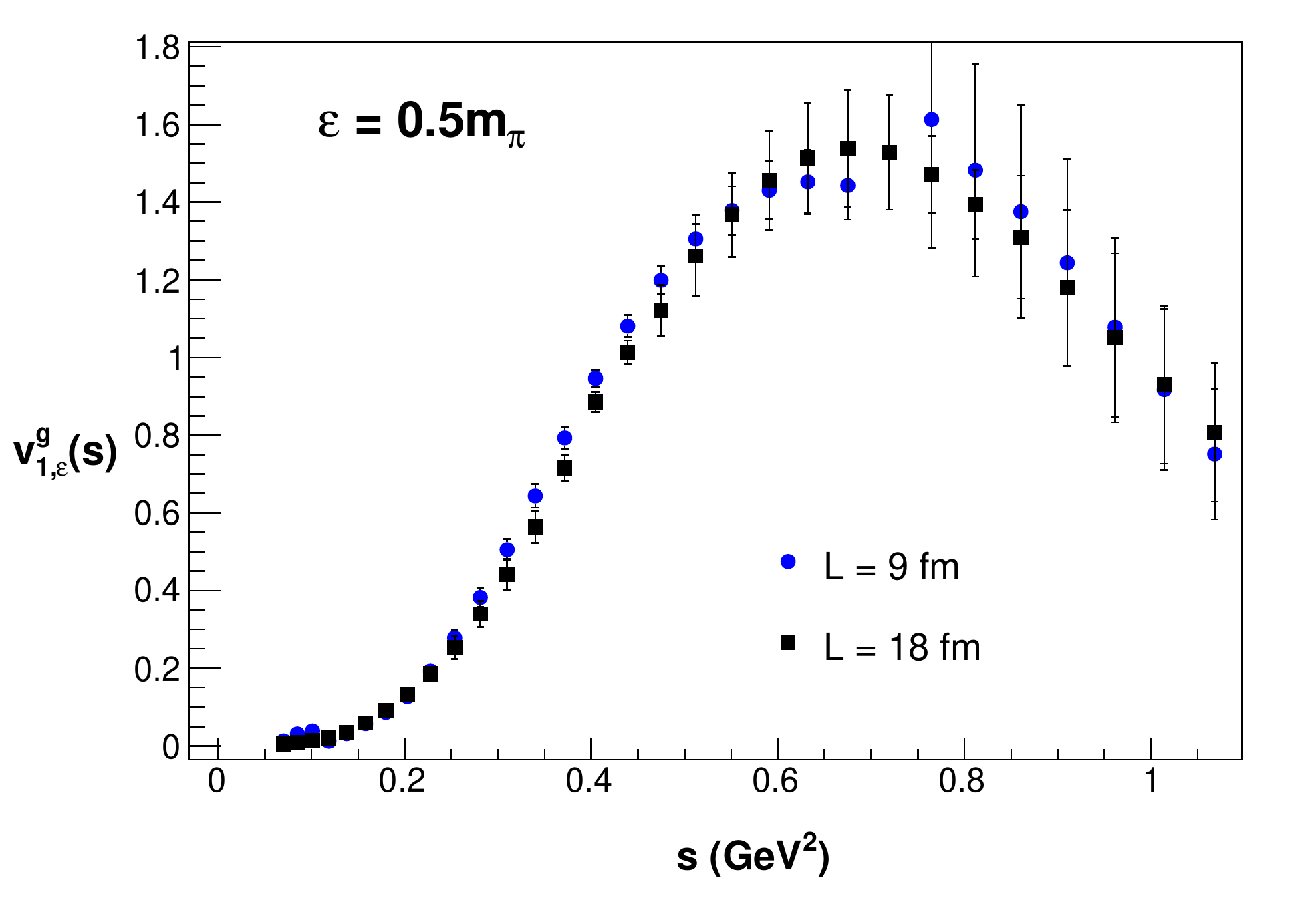}
	\includegraphics[width=0.49\textwidth]{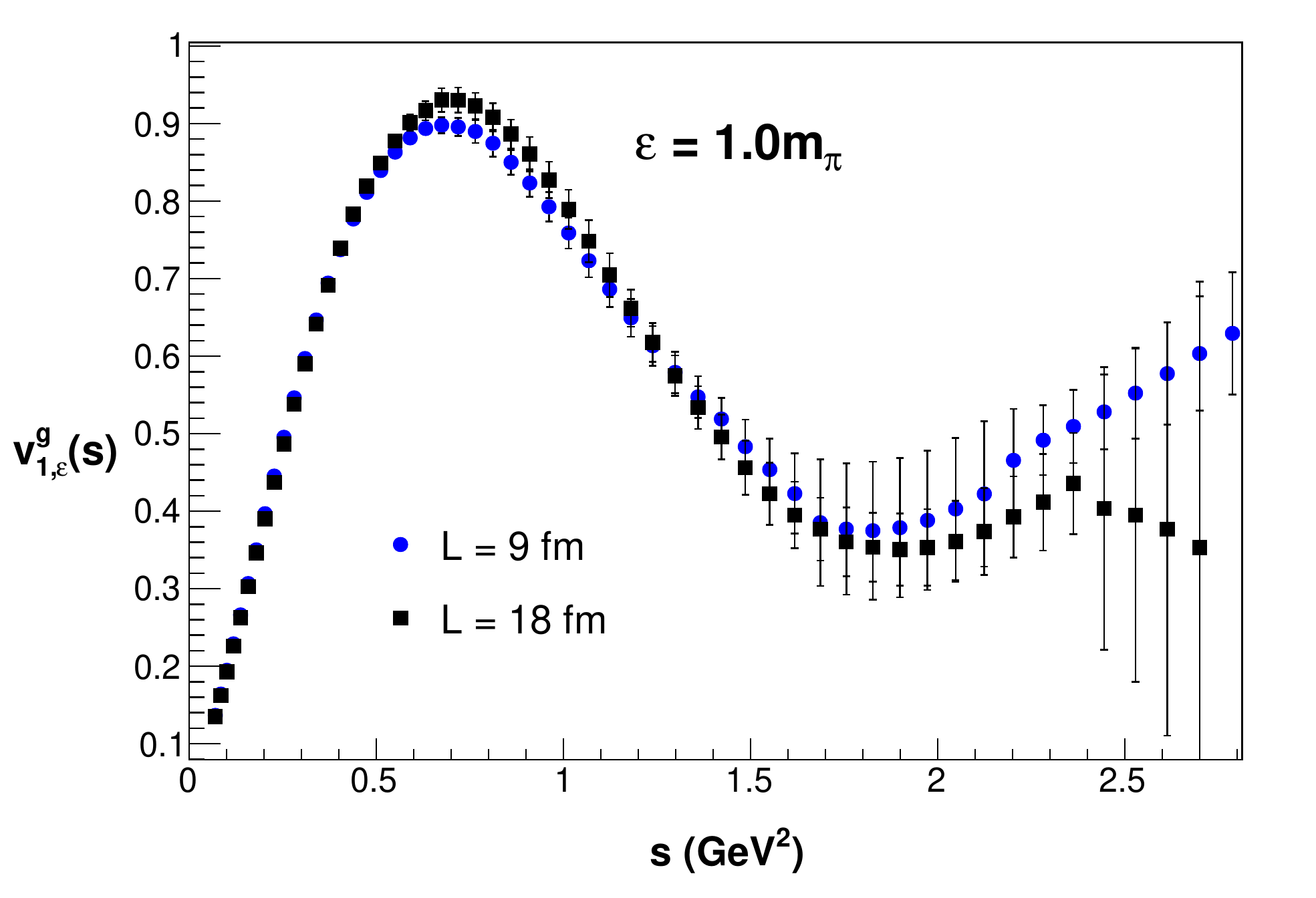}
	\caption{\label{f:fv_qcd}Finite volume effects in the reconstructed vector isovector spectral density on the two masterfield ensembles described in the text. 
	Gaussian smearing is used for a variety of energies at smearing width $\epsilon = 0.5m_{\pi}$,
	shown on the left, and $\epsilon = m_{\pi}$, shown on the right. These effects 
	are generally small apart from some mild discrepancies near $s = 0.4 \, {\rm GeV}^2$ for $\epsilon = 0.5m_{\pi}$ and $s = 0.75 {\rm GeV}^2$ for $\epsilon = m_{\pi}$. Additional smaller lattice volumes could further examine these potential finite volume effects. } 
\end{figure}

After applying the procedure discussed above for a variety of $\epsilon$ and $E$ for the gaussian kernel on each of the $L=9\,{\rm fm}$ and $18\,{\rm fm}$ ensembles, finite volume effects can be examined. This is done in Fig.~\ref{f:fv_qcd}, using $v_1(s) = 2\pi^2\rho(s)$   
for a variety of energies at two different values of the smearing width $\epsilon$. While there are possibly hints of finite-volume effects at the one-to-few sigma level at both $\epsilon$, these effects are generally under control. Additional volumes will however elucidate the situation in the future.  

We finally turn to a comparison of the reconstructed isovector vector spectral density with experiment~\cite{ALEPH:2005qgp}. For this a preliminary value of the vector current renormalization factor $Z_V$ is employed, which was presented at this conference by J. Kuhlmann. The statistical error on $Z_V$ is ignored in these preliminary results, as is the error on the lattice scale, which is crudely set by assuming $m_{\pi} = 265\, {\rm MeV}$. The results are summarized in Fig.~\ref{f:final}, and broadly resemble the experimental plot, with a narrow peak likely due to the $\rho(770)$ vector resonance followed by a slow rise due to four-particle 
states. Particularly interesting is the mild indication of this rise in the lattice QCD data, which (like in the O(3) model) show the effects of four-pion states. It should be noted that the current state-of-the-art for the finite-volume approach to lattice QCD scattering amplitudes is the numerical computation of 
(exclusive rather than inclusive) three-pion scattering amplitudes\footnote{For a review of the current status of computations of three-particle scattering amplitudes using the finite-volume approach, see the presentation by F. Romero-{L\'{o}pez} and the recent reviews in Refs.~\cite{Hansen:2019nir,Mai:2022eur}.}. 
\begin{figure}
	\includegraphics[width=0.53\textwidth]{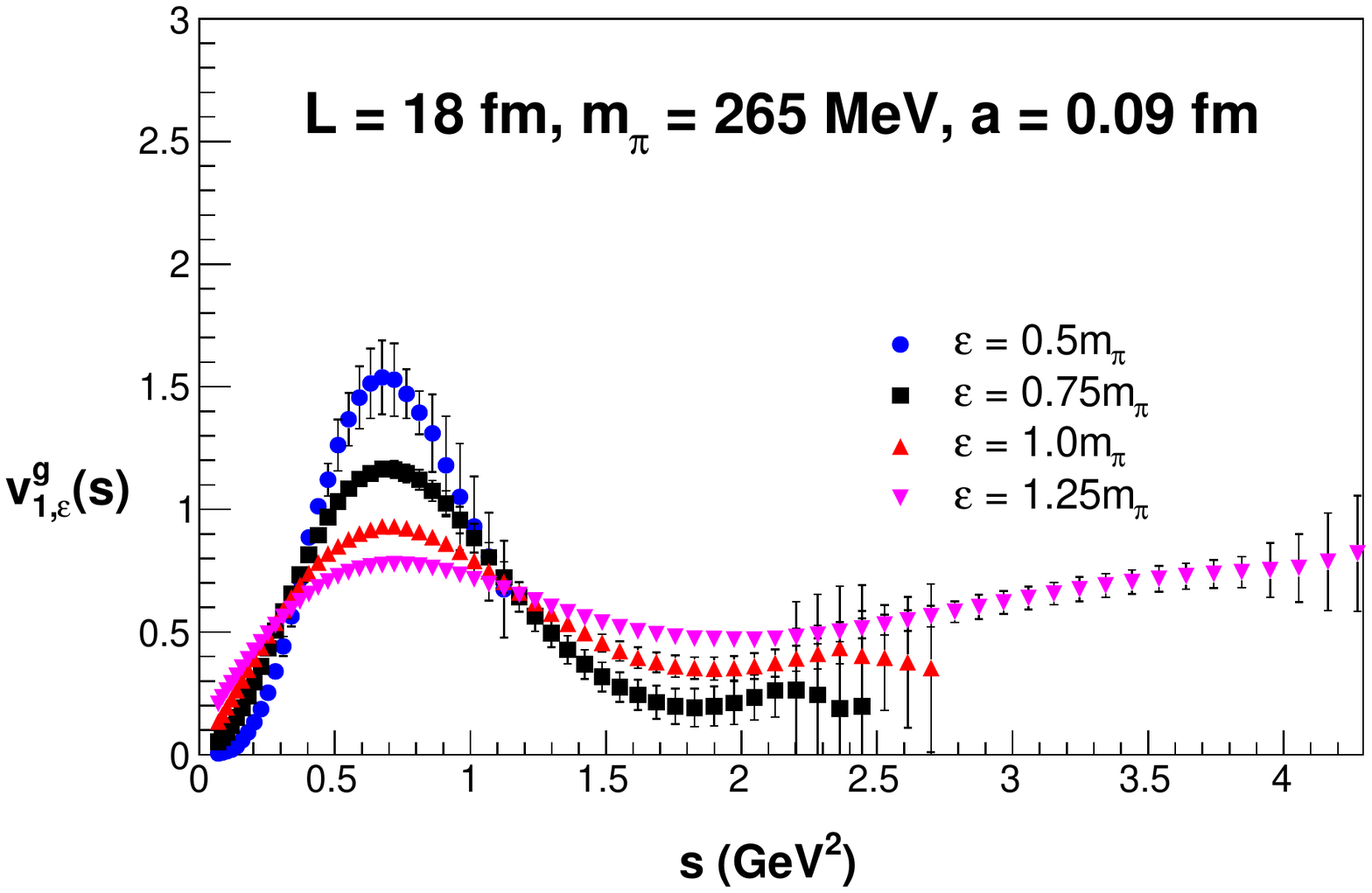}
	\includegraphics[width=0.445\textwidth]{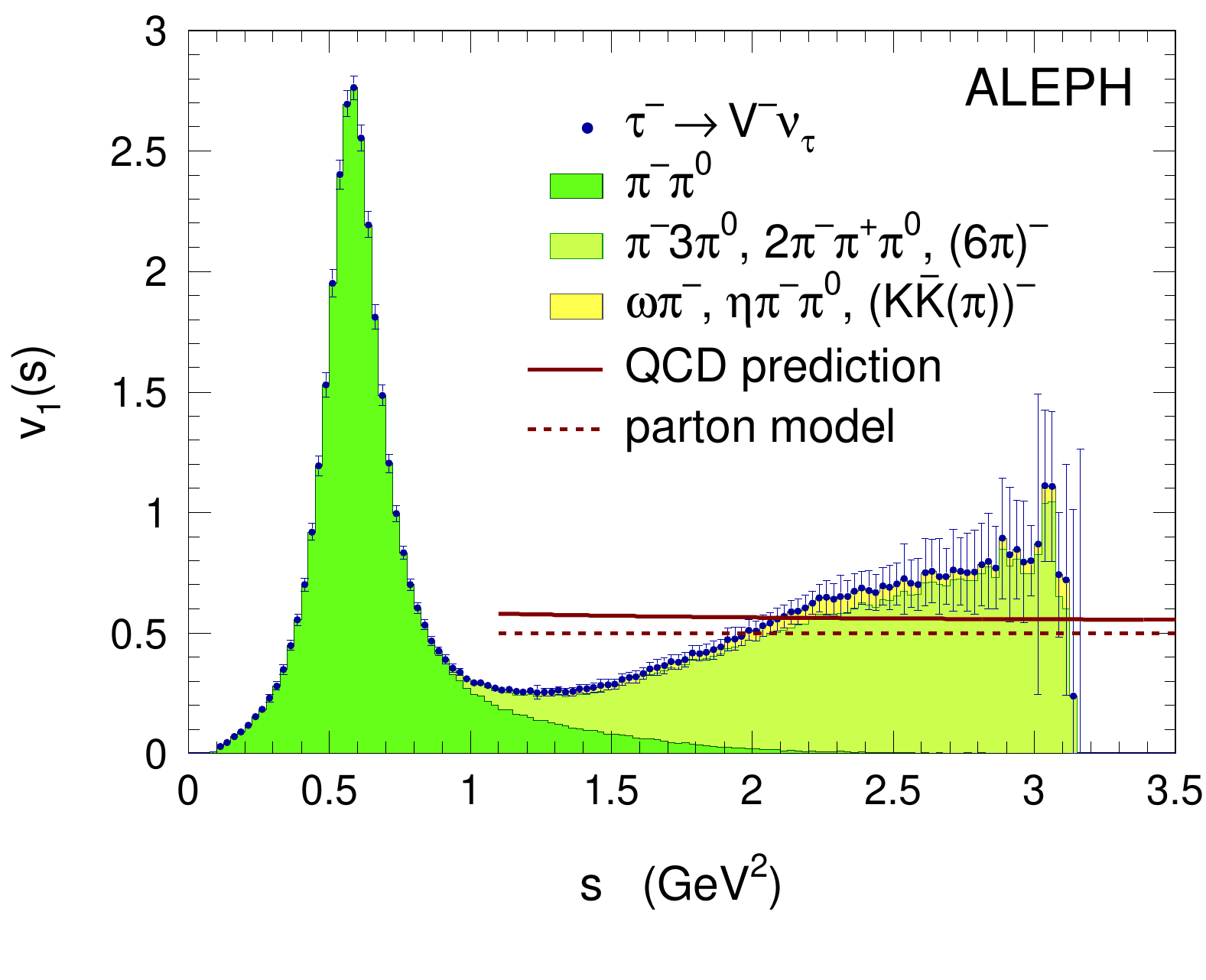}
	\caption{\label{f:final} Comparison of the lattice QCD results for the isovector vector spectral density on the larger $L/a=192$ master field ensemble discussed in the text (shown on the left), with experimental results for hadronic $\tau$-decays on the right. Statistical errors due to the scale setting and the renormalization of the vector current are not yet taken into account.}  
\end{figure}

\section{Conclusions}\label{s:conc}

Alternative techniques are required to compute phenomena arising from
many hadronic states. The spectral reconstruction of smeared spectral densities 
from euclidean correlator data not only bridges the gap between finite and 
infinite volume, but also helps to regulate the ill-posed nature of the problem.
The application discussed here is the computation of inclusive rates summed over 
all final states produced by an external current. In the two-dimensional 
O(3) model, after taking the continuum limit,  
the algorithm presented in Sec.~\ref{s:o3} (first proposed for lattice field theory in Ref.~\cite{Hansen:2019idp}) results in smeared spectral densities 
consistent with known analytic results. Spectral reconstruction algorithms 
based on the Backus-Gilbert approach~\cite{BG1, BG2} enable a precise definition of the 
smeared spectral density that has been computed, while the modification of Ref.~\cite{pt_hlt} further allows the \emph{a priori} specification of a 
desired smearing kernel. The simple linear ansatz on which these approaches are based enables the direct expression for the smearing kernel given in Eq.~\ref{e:ker}. 

Smeared spectral densities are useful not only for inclusive decay rates. An incomplete list of recent applications of the Backus-Gilbert approach includes the nucleon hadronic tensor~\cite{Liang:2019frk}, the determination of PDFs from Ioffe time data~\cite{Karpie:2019eiq}, and the photon emissivity of the quark-gluon plasma~\cite{Ce:2022fot}. These applications do not employ the algorithmic variant enabling \emph{a priori} specification of the smearing kernel, but could 
perhaps benefit from it in the future.  
This \emph{a priori} specification of the kernel 
enabled in Refs.~\cite{Hansen:2019idp,pt_hlt} is also present in 
the Chebyshev approach of Ref.~\cite{Bailas:2020qmv}, but the 
stabilizing effect of the functional $B[g]$ in Eq.~\ref{e:funcs} is naively 
not present. The work of Ref.~\cite{Barone:2022gkn} is a first step towards
comparing the two approaches. 

The advantages of the \emph{a priori} approach are 
 leveraged in the two-dimensional O(3) model to perform joint constrained $\epsilon\rightarrow 0$ extrapolations with several different kernels. The presence 
 of the narrow $\rho(770)$ peak in the isovector vector spectral density in QCD discussed in Sec.~\ref{s:qcd} complicates such an extrapolation and more work is 
 required toward an implementation. 
 A similar approach has been employed to compute inclusive decay rates in Refs.~\cite{Gambino:2022dvu,Gambino:2020crt}, and taken up by additional groups in Refs.~\cite{Kellermann:2022mms,Barone:2022gkn,Gambino:2022rai}. Work towards computing the $R$-ratio was reported in this conference~\cite{Alexandrou:2022sdj}, as well as a similar analyses of 
 the total hadronic tau decay rate~\cite{Evangelista:2023vtl,Alexandrou:2022tyn}, albeit with a wider gaussian smearing radius than 
 employed here. The interplay between the spatial extent and the smallest achievable smearing width requires further study.  
 Furthermore, the \emph{a priori} approach of
 Ref.~\cite{Bailas:2020qmv} led to the 
direct computation of the Borel transform of a current-current correlator
 required for the Shifman-Vainshtein-Zakharov sum rule in Ref.~\cite{Ishikawa:2021txe}, 
possibly opening the door for additional interaction between 
lattice QCD and QCD sum rules. Another interesting application is pursued in 
Ref.~\cite{DelDebbio:2022qgu} in which
fits to smeared spectral densities are considered as an alternative 
to `standard' spectroscopy. 
Additional applications could appear in the future. The \emph{a priori} approach 
in principle enables the computation of exclusive 
scattering amplitudes using Refs.~\cite{Bulava:2019kbi,Bruno:2020kyl}, while the formalism for 
inclusive rates was developed already in Refs.~\cite{Hansen:2017mnd,Fukaya:2020wpp}.



\bibliography{lattice}

\end{document}